\documentclass[aps,twocolumn,showpacs,groupedaddress,preprintnumbers]{revtex4}

\usepackage{epsf}
\usepackage{graphicx}
\usepackage{dcolumn}
\usepackage{bm}

\begin{document}
\bibliographystyle{apsrev}


 \flushbottom
 \renewcommand{\textfraction}{0}
 \renewcommand{\topfraction}{1}
 \renewcommand{\bottomfraction}{1}

\newcommand{\beq}{\begin{equation}}
\newcommand{\dd}{\partial}
\newcommand{\eeq}{\end{equation}}
\newcommand{\bea}{\begin{eqnarray}}
\newcommand{\eea}{\end{eqnarray}}



\preprint{UCLA/04/TEP/38}

\title{Effects of Gravity and Finite Temperature on the Decay of the False
Vacuum}

\author{Lee C. Loveridge$^1$}
\affiliation{$^1$Department of Physics and Astronomy, UCLA, Los Angeles, CA
90095-1547}

\begin{abstract}
\vspace{0.1cm}

I have calculated the exponential suppression factor in the decay rate of
the false vacuum (per unit volume) for a real scalar field at finite
temperature, in the presence of gravity, in the thin-wall approximation.
Temperatures are assumed to be much greater than the inverse of the
nucleation radius.  The value of a local minimum of the scalar potential
is arbitrary.  Thus, both true and false vacuua may have arbitrary
cosmological constants.

\end{abstract}

\pacs{11.10.-z, 11.10.Wx, 11.27.+d}

\maketitle

\section{introduction}
Vacuum decay occurs when the potential for some scalar field has a local
minimum which is not the global minimum.  If the field starts out in the
local minimum, the resulting ``false vacuum'' can remain metastable for
some time, until it decays by tunneling.  The decay of the false vacuum has
been studied by many authors.  For example, the exponential decay factor was 
first found by \cite{Kobzarev:1974cp} then later more rigorously by 
\cite{Coleman:1977py}.  The prefactor which contains some of the basic 
quantum mechanical modifications to the decay probability was found for a 
single scalar field in \cite{Callan:1977pt} and for spontaneous symmetry
breaking and internal degrees of freedom in \cite{Kusenko:1995bw} and
\cite{Kusenko:1996bv} respectively.

This effect is modified in the presence of gravity
\cite{Coleman:1980aw}, \cite{Parke:1982pm} or at finite temperature
\cite{Linde:zj}.  As demonstrated in Ref.~\cite{Parke:1982pm}, gravity
enhances the decay process when the average of the two minima is positive
and diminishes it if both minima are negative.  If the false minimum is
positive, but the average is negative, the behavior is slightly more
complex: the vacuum decay rate decreases if the gravitational effects are small
and increases if the gravitational effects become large.

Reference \cite{Linde:zj}, gives an excellent explanation of how the decay
rate changes at finite temperatures.  First, the potential is changed, second
the action integral is taken over a periodic time interval of period 
$\frac 1 T$.  
Thus, if the bubble radius is much larger than this period, we must change from
a bubble that is a 4-dimensional sphere to one that is a 4-dimensional cylinder
and whose cross section is a 3-dimensional sphere.  
The transition between these two limits is hinted at
but not fully explored in the reference.  I am not aware of any work which 
explores this more thoroughly, nor will I attempt to here.

In this paper I will combine these procedures to get a description of the 
decay rate of the false vacuum in the presence of gravity and at a finite 
temperature.  Some work in this direction has been done before 
\cite{Su:1982un}.  However, they considered only the change in the potential
caused by the finite temperatures, and did not consider the change in the 
action integral.

Other exotic decay possibilities have also been explored.  \cite{Lee:1987qc} 
explores the possibility that in an expanding universe, the true vacuum
may tunnel back to a false vacuum. \cite{Lee:1985uv} and \cite{Lee:1986sa}
explore the possibility of tunneling decays in the absence of minima and 
barriers in the potential in flat and curved space respectively.  
\cite{Abbott:1984qf} shows how gravitational stabilization could explain
the smallness of the cosmological constant.  These ideas in conjunction with 
the results of this paper may provide interesting avenues for further study.

\section{Calculation of Bounce at Finite Temperature}

Let us consider a scalar field with a potential that has at least two 
non-degenerate local minima at a given temperature.  We will call $\phi$ in
the greater (false) vacuum state $\phi_+$ and the potential at this point
$U_+ \equiv U(\phi_+,T)$.  Similarly, $\phi$ at the lesser (possibly true) 
vacuum state is $\phi_-$ and the potential there is $U_-\equiv U(\phi_-, T)$.

Vacuum decay will begin when a bubble of the lesser state nucleates in the 
greater state.  The nucleation rate per unit volume is 
$A \exp [-B/\hbar](1+O(\hbar))$.  B is the action of the bounce.  
(Or more accurately the difference between the action of $\phi_+$ and
the action of the bounce.)  The bounce, $\phi_b$ is a choice of $\phi$ which 
\begin{enumerate}
\item {solves the euclidean equations of motion,}
\item {transverses the barrier between the two vacuua, and}
\item {minimizes the action for these two criteria.}
\end{enumerate}

In the absence of gravity or finite temperatures, the bounce is 
spherically symmetric in four euclidean space-time dimensions and is described
by a bubble of lesser (true) vacuum around $\rho=0$ (where $\rho$ is the 
radial variable), a thin shell at $\bar{\rho}$ where the value of $\phi$
changes from $\phi_-$ to $\phi_+$ and then the outside of the bubble is simply
$\phi_+$.  
 
Following Linde \cite{Linde:zj}, we notice that
because we are considering a system in thermal equilibrium, the action integral
must be periodic in imaginary time with a period $\beta = 1/T$.   
If the radius of the nucleating bubble is much smaller than $\beta$ then this
change is irrelevant as the appropriate periodic system is just a series
of such bubbles a distance $\beta$ apart.  However, if the temperature is 
much higher so that $\beta$ is very small compared with the radius of the
nucleating bubble, then the bounce corresponds to a cylinder.  That is to say
a three sphere in the spatial directions and constant in time.  The integration
over time is then simply multiplication by $\beta=1/T$.

In the absence of gravity it can be proved that the solution to the bounce
is spherically symmetric at zero temperature in 4 Euclidean dimensions, and
is a 4-cylinder (spherically symmetric in three orthogonal dimensions) at
high temperature.  In the presence of gravity this is only an assumption,
so these results are only a lower bound on the nucleation rate.  However,
the assumption seems reasonable and no counter-example is known.

In evaluating the effects of gravity I will follow \cite{Coleman:1980aw}, and
\cite{Parke:1982pm}.  
I begin with the action for a scalar field including gravity
\begin{equation}
S=\int d^4 x \sqrt{-g} \left[ \frac 1 2 g^{\mu \nu} \partial_\mu \phi
\partial_\nu \phi - U(\phi,T) - (16 \pi G)^{-1}R \right],
\end{equation}
and a metric of the form
\begin{equation}
(ds)^2=(d\tau)^2 + (d\xi)^2 + \rho(\xi)^2(d \Omega)^2.
\end{equation}

Computing the Euclidean equations of motion is now straightforward.  
The results follow.
By varying with respect to $\phi$ we find the equation
\begin{equation}
\phi''+ 2 \frac {\rho'} {\rho} \phi' = \frac {dU} {d\phi}.
\label{eqmotion}
\end{equation}
Varying with respect to the metric yields the Einstein Equation
\begin{equation}
G_{\mu \nu} = \kappa T_{\mu \nu},
\end{equation}
the $\xi \xi$ component of which becomes
\begin{equation}
\rho'^2=1+\kappa \rho^2 (\frac 1 2 \phi'^2 - U).
\label{einstein2}
\end{equation}
Finally the action can be rewritten as
\begin{equation}
S_E=\frac {4 \pi} T \int d\xi \left[\rho^2 
          \left(\frac 1 2 \phi'^2 +U \right) + \frac 1 {\kappa}
          \left(2 \rho \rho''+ \rho'^2 - 1 \right) \right].
\label{action2}
\end{equation}

At this point Coleman \cite{Coleman:1980aw} finds $\phi$ and $\rho$ 
in terms of $\xi$.  This calculation is important for determining the validity
of the thin-wall approximation.  I will not repeat this calculation as it is 
unnecessary to understand the derivation of the bounce action. 
The calculation is the same as in the reference.

We can now simplify the action (\ref{action2}) by using integration by parts to
eliminate the second derivative term, and equation (\ref{einstein2})
 to eliminate $\rho'$ which yields
\begin{equation}
S_E=\frac {8 \pi} T \int d\xi \left( \rho^2 U - \frac 1 {\kappa} \right)
+{\rm surface \hspace{5 pts} terms.}
\end{equation}

The bounce action is then 
\beq
B \equiv S_E(\phi_b)-S_E(\phi_+).
\eeq
This can be divided into three parts.  Outside the wall, the wall itself,
and inside the wall.

Because the bounce solution and greater vacuum are identical outside of the 
wall, there is no contribution to the bounce from outside the wall.  
(This is also why surface terms can be safely neglected.)
\begin{equation}
B_{outside}=0
\end{equation}
At the wall it is useful to define $U_0(\phi,T)$ such that 
$U_0(\phi,T)=U(\phi,T)+O(U_+-U_-)$, $U_0(\phi_+,T)=U_0(\phi_-,T)$, and
$dU_0/d\phi = 0$ at both $\phi_+$ and $\phi_-$.  
We can now approximate $\rho$ as $\bar{\rho}$ and $U(\phi,T)$ as 
$U_0(\phi,T)$ to get
\begin{equation}
B_{wall}=\frac {8 \pi} T \bar{\rho}^2 \int d\xi
         \left[ U_0(\phi,T)-U_0(\phi_+,T) \right]
        =\frac {4 \pi} T \bar{\rho}^2 S_1.
\end{equation} 

Inside the wall, $\phi$ is constant so that from equation (\ref{einstein2})
we get
\begin{equation}
d\xi=d\rho(1-\kappa \rho^2 U)^{-1/2},
\end{equation}
so that
\begin{eqnarray}
B_{in}&=&-\frac {8 \pi} {T \kappa} \int_0^{\bar{\rho}} d \rho 
\left\{\left[1 - \kappa \rho^2 U(\phi_-)\right]^{\frac 1 2} 
- (\phi_- \to \phi_+)\right\}  \label{inbounce} \\
&=& -\frac {4 \pi} {T \kappa } \left\{
\left[\frac {\arcsin{\sqrt{\kappa |U(\phi_-)}|\bar{\rho}}} 
{\sqrt{\kappa |U(\phi_-)|}}
+\bar{\rho}\sqrt{1-\kappa U(\phi_-)\bar{\rho}^2} \right] \right. \nonumber \\
&& \hspace{2 cm}  -(\phi_- \to \phi_+) \Bigg\}.
\nonumber 
\end{eqnarray} 
I should note that when $U<0$ the $\arcsin$ must be changed to an 
inverse hyperbolic $\sin$.  (This can also be accomplished by simply dropping 
the absolute value signs.)  Otherwise the equation is unchanged.

\section{Simple Cases}
While I will, in the next section, 
explore the general case at finite temperature (as
Parke \cite{Parke:1982pm} did for the zero temperature case), it is 
illustrative to begin by considering the relatively simple cases
given in \cite{Coleman:1980aw}.  It is useful to understand the
simpler solutions first as a reference and check for the more general solution.
It is also worth noting that in finding the extrema of the bounce solutions
it is much easier to work from the integral form of equation (\ref{inbounce}).

\subsection{Null True Vacuum}
We begin with the case where the true vacuum is $0$ (null) and the false
vacuum is small and positive.
\begin{equation}
U(\phi_+,T)=\epsilon, \hspace{1 cm} U(\phi_-,T)=0.
\end{equation}
In this case we find
\begin{equation}
\bar{\rho}=\frac {2 S_1} {\epsilon + \kappa S_1^2}
=\frac{\bar{\rho}_0} {1 + (\frac {\bar{\rho}_0} {2 \Lambda})^2}
\end{equation}
Where $\bar{\rho}_0 = \frac {2 S_1} \epsilon$, in agreement with Linde's
work, and $\Lambda = (\kappa \epsilon)^{-1/2}$.

The bounce action can not be put in quite as simple a form as in Coleman's 
paper, but it can be written as
\begin{eqnarray}
B&=&\frac {4 \pi} T {\Lambda}^3 \epsilon \left[ 
\arcsin \left(\frac {2\alpha} {1+\alpha^2} \right) 
-\left(\frac {2\alpha} {1+\alpha^2} \right)  \right] \label{nulltrue}
\\ \nonumber
&=&\frac {4 \pi} T {\Lambda}^3 \epsilon \left[ 
\arccos \left(\frac {1-\alpha^2} {1+\alpha^2} \right) 
-\left(\frac {2\alpha} {1+\alpha^2} \right)  \right]
\end{eqnarray}

Where $\alpha = \frac {\bar{\rho}_0} {2 \Lambda}$.  We can easily verify that
this gives Linde's result as $\Lambda \to \infty$.  The need for the final 
change to an inverse cosine is apparent when you realize that the first
form traces both the domain and range of the inverse sine twice yielding
incorrect results.  The second form traces the entire domain and range
of the inverse cosine once giving correct answers.

\subsection{Null False Vacuum}
The opposite case where the false vacuum is $0$ (null) and the true vacuum is
small and negative can also be worked out rather simply.
\begin{equation}
U(\phi_+)=0, \hspace{1 cm} U(\phi_-)=-\epsilon.
\end{equation}
We find
\begin{equation}
\bar{\rho}=\frac {2 S_1} {\epsilon - \kappa S_1^2}
=\frac{\bar{\rho}_0} {1 - (\frac {\bar{\rho}_0} {2 \Lambda})^2},
\end{equation}
and the bounce is
\begin{eqnarray}
B&=&\frac {4 \pi} T {\Lambda}^3 \epsilon \left[ 
\left(\frac {2\alpha} {1-\alpha^2} \right)  
-{\sinh}^{-1} \left(\frac {2\alpha} {1-\alpha^2} \right) 
\right] \label{nullfalse}
\\ \nonumber
&=&\frac {4 \pi} T {\Lambda}^3 \epsilon \left[ 
\left(\frac {2\alpha} {1-\alpha^2} \right)  
-{\cosh}^{-1} \left(\frac {1+\alpha^2} {1-\alpha^2} \right) 
\right].
\end{eqnarray}

As in the zero temperature case, the stabilizing effect is still present.  That
is to say that if  $\left(\frac {\bar{\rho}_0} {2 \Lambda}\right)^2 >1$
then the new vacuum state is not large enough to hold the bubble and a decay
can never nucleate.

\section{The General Case}
In general neither the greater (false) vacuum $U_+$ nor the 
lesser (true) vacuum $U_-$ is $0$.  
When we allow for this the equations get far more complicated though still
reasonably tractable.  We find that
\begin{eqnarray}
\bar{\rho}^2&=&\frac {4 {S_1}^2} 
{\kappa^2 {S_1}^4 + 2 {S_1}^2 \kappa (U_+ + U_-) + (U_- - U_+)^2} \nonumber \\
&=&\frac {{\bar{\rho}_0}^2} {1+2\left(\frac {\bar{\rho}_0} {2 \lambda}\right)^2
+\left(\frac {\bar{\rho}_0} {2 \Lambda}\right)^4}.
\end{eqnarray}
As in the previous section, $\bar{\rho}_0$ is the critical radius in the 
absence of gravity
\begin{equation}
\bar{\rho}_0=\frac {2 S_1} {U_+-U_-}.
\end{equation}
Also
\begin{equation}
\lambda^2=[\kappa (U_+ + U_-)]^{-1},
\end{equation}
and
\begin{equation}
\Lambda^2=[\kappa (U_+ - U_-)]^{-1}.
\end{equation}

Now when we add the parts of the bounce from the wall and the interior
of the bubble, we find that the simple square-root terms cancel leaving only
the $\arcsin$ terms so that
\begin{equation}
B=\frac {4 \pi} {\kappa T}
\left[\frac {\arcsin {\sqrt{\kappa U_+} \bar{\rho}}} {\sqrt{\kappa U_+}}
-\frac {\arcsin {\sqrt{\kappa U_-} \bar{\rho}}} {\sqrt{\kappa U_-}} \right]
\end{equation}
To check this we take the limit in which $\kappa \to 0$ (small gravity) and
find
\begin{equation}
B_0=\frac {2 \pi} {3T} {\bar{\rho}_0}^3 \epsilon
= \frac {16 \pi {S_1}^3} {3 T \epsilon^2}
\end{equation}
in agreement with \cite{Linde:zj}.

As in \cite{Parke:1982pm} it is useful to separate out the zero gravity 
portion of the bounce and write
\begin{equation}
B=B_0 r[({\bar{\rho}_0}/{2 \Lambda})^2, \Lambda^2/\lambda^2]
\end{equation}

There are many ways to write $r(x,y)$.  These are the two which I find most 
clear and illuminating.

\begin{eqnarray}
r(x,y)&=&\frac 3 {2 x^{3/2}} 
\int_0^x \frac {z^{1/2}} {1+2yz+z^2} dz \label{intform}\\
&=&\frac 3 { 2 \sqrt{2} x^{3/2}} \left[ \frac {1} {\sqrt{y+1}} 
\arccos { \frac {1-x} {\sqrt{1+2 y x +x^2}}} \right. \label{arcsinform} \\
&& \left. \hspace{1.5 cm} - \frac {1} {\sqrt{y-1}} 
\arccos { \frac {1+x} {\sqrt{1+2 y x+x^2}}} \right]
\nonumber
\end{eqnarray}

Both forms are relatively simple, though for $y<1$ equation 
(\ref{arcsinform}) is not explicitly real.  (That is to say two factors of
$i$ arise and cancel.)  The integral form (\ref{intform}) is always explicitly
real, and deals with the limits as $y\to \pm 1$ and $x\to 0$ 
more easily.  However, actually evaluating the integral is very difficult 
to do in closed form.  So both forms are useful.  Also, if $y=\pm 1$ and
$x=\alpha^2$ both equations (\ref{intform}) and (\ref{arcsinform}) reduce
to forms consistent with equations (\ref{nulltrue}) and (\ref{nullfalse}).

\begin{figure}
\begin{center}
\includegraphics[height=6cm]{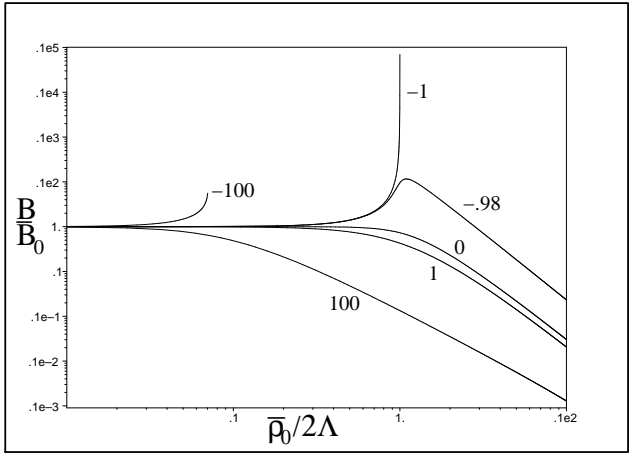}%
\end{center}
\caption{The ratio $B/B_0(=r)$ as a function of $(\bar{\rho}_0/2\Lambda)$ for
various values of $\Lambda^2/\lambda^2$.  The number next to each line gives
the value of $\Lambda^2/\lambda^2$ for that specific line.}
\label{bounce2fig}
\end{figure}

Figure \ref{bounce2fig} shows $B/{B_0}$ ($r$) as a function of 
$\alpha=\frac {\bar{\rho}_0} {2 \Lambda}$ for various values of
$\Lambda^2/\lambda^2$
It is clear from the figure that all the same basic features exist as in
the zero temperature case.
\begin{enumerate}
\item For $\Lambda^2/\lambda^2>0$ (which  implies
$U_+ + U_- >0$), gravity lessens the bounce action and thus
increases the rate of decay.  It is interesting to consider two limits of
this case.
\begin{enumerate}
\item If $\lambda^2/\Lambda^2 \ll (\bar{\rho}_0/2 \Lambda)^2 
\ll \Lambda^2/\lambda^2$, then
\begin{eqnarray}
r[(\bar{\rho}_0/2 \Lambda)^2,\Lambda^2/\lambda^2]&=&
3/\left\{2\left[(\bar{\rho}_0/2 \Lambda)^2 \Lambda^2/\lambda^2 \right]\right\}
 \nonumber \\
&=& 6 \lambda^2 / {\bar{\rho}_0}^2
\end{eqnarray}
\item $(\bar{\rho}_0/2 \Lambda)^2 \gg 1$ and 
$(\bar{\rho}_0/2 \Lambda)^2 \gg \Lambda^2/\lambda^2$ then 
\begin{eqnarray} 
r[(\bar{\rho}_0/2 \Lambda)^2,\Lambda^2/\lambda^2]&=&
\frac {3 \pi} {2 \sqrt{2(\Lambda^2/\lambda^2 +1)} (\bar{\rho}_0/2 \Lambda)^3}
\nonumber \\
&=&
\frac {3 \pi} 4 \sqrt{\frac {U_+-U_-} {U_+}} / 
{\left( \frac {\bar{\rho}_0} {2 \Lambda} \right)}^3.
\end{eqnarray}
\end{enumerate}
(The last equation still applies in the next situation.)  So we see
that the system will change from quadratic to cubic falloff around
\beq
(\bar{\rho}_0/2\Lambda) \cdot \lambda/\Lambda =1.
\eeq
\item For $-1<\Lambda^2/\lambda^2<0$ (which implies $U_+>0$, $U_+ + U_- <0$)
gravity increases the bounce action for small
values of $(\bar{\rho}_0/(2 \Lambda))$, but the bounce action reaches a maximum
and for large values of $(\bar{\rho}_0/(2 \Lambda))$ the action falls off as
it does for positive $\Lambda^2/\lambda^2$.  The maximum action will occur when
the value of $\bar{\rho}_0/2 \Lambda$ is slightly larger than the value
which minimizes
$1+2\Lambda^2/\lambda^2 (\bar{\rho}_0/(2 \Lambda))^2 
               + (\bar{\rho}_0/(2 \Lambda))^4$.
\item For $\Lambda^2/\lambda^2\leq-1$ (which implies $U_+,U_- < 0$) 
gravity increases the bounce action 
which goes to infinity (completely stabilizing the vacuum) when 
$1+2\Lambda^2/\lambda^2 (\bar{\rho}_0/(2 \Lambda))^2 
               + (\bar{\rho}_0/(2 \Lambda))^4=0$.  In other words,
the greater vacuum is stable if 
\beq
(\bar{\rho}_0/2 \Lambda)^2 \geq -\Lambda^2/\lambda^2 - 
\sqrt{\Lambda^4/\lambda^4-1},
\eeq
which is the same as when
\beq
S_1 \sqrt{8 \pi G} \geq \sqrt{-U_-} - \sqrt{-U_+}.
\eeq

\end{enumerate}

\section{Range of Validity}
These results are valid under the same conditions as  in \cite{Parke:1982pm}.
That is $B \gg 1$ so that the semi-classical treatment is valid, and all
length scales $\bar{\rho}_0$, $|\lambda|$, and $\Lambda$ must be much greater
than the thickness of the wall.  

The most restrictive of these is likely to be
$|\lambda|$ because while we set the difference between $U_+$ and $U_-$ to
be small in all cases, we have never yet required that these potentials
themselves be small. (In the absence of gravity the value of the potentials
themselves is irrelevant.  Only the difference mattered.)

The thickness of the wall is determined by an equation of the approximate form
\begin{eqnarray}
\Delta \xi&=&\int_{\frac{(\phi_+ + 3\phi_-)} 4}^{\frac{(3\phi_+ + \phi_-)} 4} 
d\phi
\left\{2 \left[U_0(\phi),T-U_0(\phi_{\pm},T)\right]\right\}^{-\frac 1 2} 
\nonumber \\
&\sim& \frac {\Delta \phi} {(U_{max}-U_{min})^{1/2}}
\end{eqnarray}
\cite{Coleman:1980aw}.  From this we find that the restriction that
$|\lambda|$ is much greater than the thickness of the wall can be given
approximately by
\beq
\frac {U_{max}-U_{min}} {|U_{min}|} >> (\Delta \phi)^2 \kappa.
\eeq
As long as gravity is weak and the difference between $\phi_+$ and $\phi_-$ is
not to great, this should easily be satisfied by a moderately high 
barrier.  As gravity or the difference between the values of $\phi$ increases
the barrier must be significantly greater than the actual values of the
two minima in the potential.

We should also note that to work in the finite temperature case we assumed
$T \gg \frac 1 {\bar{\rho}}$.  In this case the bounce action will always be 
significantly less than in the zero temperature case by a factor of about
\beq
B(T)/B(0) \sim \frac 1 {\bar {\rho} T} \sim \frac {\epsilon} {S_1 T}.
\eeq 
(Because the potential also changes at finite temperature, there is no reason
to improve this estimate.)

\section{Conclusion}
We have seen in this paper that many of the basic features of vacuum decay in
the presence of gravity are unchanged at finite temperature.  Gravity still
enhances the decay of positive energy vacua and suppresses the decay of 
negative energy vacua, and there is still a middle region of decay from a 
positive energy vacuum to a negative energy vacuum (whose energy has a greater
magnitude) which is suppressed when gravitational effects are small and 
enhanced as they increase.  Also we see that thermal effects increase the rate
of decay just as they did in flat space.  There are many interesting avenues
for further study including the possibility that the true vacuum may actually
''decay'' to the false vacuum, the effects of such tunneling in very flat 
potentials, and potential implications for cosmology.

I would like to thank Alexander Kusenko for many helpful discussions and ideas.
This work was supported in part by the US Department of Energy grant
DE-FG03-91ER40662.


\end{document}